\documentclass[twocolumn,showpacs,floatfix,aps,pra,showkeys]{revtex4}

\usepackage[dvips]{graphicx,color}
\usepackage{dcolumn}
\usepackage{bm}
\usepackage{amsmath, amssymb, mathrsfs}
\newcommand{\kk}{\mathbf{k}}
\newcommand{\rr}{\mathbf{r}}
\newcommand{\RR}{\mathbf{R}}
\newcommand{\CC}{\mathbf{C}}

\newcommand{\be}{\begin{equation}}
\newcommand{\ee}{\end{equation}}
\newcommand{\bea}{\begin{eqnarray}}
\newcommand{\eea}{\end{eqnarray}}

\begin{document}
\title{Four-body Efimov effect}

\author{Yvan Castin${}^1$, Christophe Mora${}^2$, Ludovic Pricoupenko${}^3$}
\affiliation{Laboratoires ${}^1$Kastler Brossel and ${}^2$Pierre Aigrain, 
\'{E}cole Normale Sup\'{e}rieure and CNRS, UPMC${}^1$  and Paris 7 Diderot${}^2$, 24 rue Lhomond, 75231 Paris, France \\
${}^3$Laboratoire de Physique Th\'eorique de la Mati\`ere Condens\'ee,
UPMC and CNRS, 4 place Jussieu, 75005 Paris, France}
\date{\today}

\begin{abstract}
We study three same spin state fermions of mass $M$ interacting with a distinguishable 
particle of mass $m$ in the unitary limit 
where the interaction has a zero range and an infinite $s$-wave scattering length.
We predict an interval of mass ratio $13.384 < M/m < 13.607$ where there exists
a purely four-body Efimov effect, leading to the occurrence of weakly bound tetramers 
without Efimov trimers.
\end{abstract}



\maketitle

In a system of interacting particles, the unitary limit corresponds to a zero range $s$-wave interaction
with infinite scattering length \cite{revue_efimov}. In particular, this excludes any finite energy two-body bound state.
Interestingly, in the three-body problem, the Efimov effect may take place \cite{Efimov}, leading to the occurrence
of an infinite number of three-body bound states, with an accumulation point in the spectrum at zero energy.
This effect occurs in a variety of situations, the historical one being the case of three bosons, 
as recently studied in a series of remarkable experiments with cold atoms close to a Feshbach resonance \cite{manips_Efimov}. 
It can also occur in a system of two same spin state fermions of mass $M$ and a particle of another species
of mass $m$, in which case the fermions only interact with the third particle, with an infinite $s$-wave
scattering length:
An infinite number of arbitrarily weakly bound trimers then appears
in this $2+1$ fermionic problem 
if the mass ratio $\alpha=M/m$ is larger than $\alpha_c(2;1) \simeq 13.607$ \cite{Efimov}.

The four-body problem has
recently attracted a lot of interest \cite{quatre_corps}.
The question of the existence of a four-body Efimov effect
is however to our knowledge still open.
We give a positive answer to this question, by investigating 
the $3+1$ fermionic problem in the unitary limit. We explicitly solve
Schr\"odinger's equation in the zero range model \cite{Efimov} and we determine the critical mass ratio
to have a purely four-body Efimov effect in this system, that is without Efimov trimers.

In the zero-range model, the Hamiltonian reduces to a non-interacting form,
here in free space
\be
\label{eq:hamil}
H = \sum_{i=1}^{4} -\frac{\hbar^2}{2m_i} \Delta_{\rr_i},
\ee
with $m_1=m_2=m_3=M$ and $m_4=m$.
The interactions are indeed replaced by contact conditions on the wavefunction,
$\psi(\rr_1,\rr_2,\rr_3,\rr_4)$, where $\rr_i$, $i=1,2,3$ is the position of a fermion and $\rr_4$ is the position
of the other species particle: At the unitary limit, for $i=1,2,3$, 
there exist functions $A_i$ such that
\be
\label{eq:BP}
\psi(\rr_1,\rr_2,\rr_3,\rr_4) = \frac{A_i(\RR_{i4};(\rr_k)_{k\neq i,4})}{|\rr_i-\rr_4|} + O(|\rr_i-\rr_4|)
\ee
when $\rr_i$ tends to $\rr_4$ for a fixed value of the $i$-4 centroid $\RR_{i4}\equiv (M \rr_i+m \rr_4)/(m+M)$
different from the positions of the remaining particles $\rr_k$, $k\neq i,4$.
The wavefunction is also subject to the fermionic exchange symmetry with respect to the first three variables
$\rr_i$, $i=1,2,3$.

In what follows, we shall assume that there is no three-body Efimov effect, 
a condition that is satisfied by imposing $M/m < \alpha_c(2;1) \simeq 13.607$.
The eigenvalue problem $H\psi=E\psi$ with the contact conditions in Eq.(\ref{eq:BP}) is then 
separable in hyperspherical coordinates \cite{Werner2006}.
After having separated out the center of mass $\CC$ of the system,
one introduces the hyperradius $R=\left[\sum_{i=1}^4 m_i (\rr_i-\CC)^2/\bar{m}\right]^{1/2}$,
with $\bar{m}=(3M+m)/4$ the average mass, and a set of here $8$ hyperangles
$\Omega$ whose expression is not required. For a center of mass at rest, the wavefunction
may be taken of the form
\be
\label{eq:separe}
\psi(\rr_1,\rr_2,\rr_3,\rr_4)= R^{-7/2} F(R) f(\Omega).
\ee
$f(\Omega)$ is given by the solution of a Laplacian eigenvalue problem on the unit sphere of dimension $8$,
which is non trivial because of the contact conditions.  On the contrary, the hyperradial part $F$ is not
directly affected by the contact conditions, due in particular to their invariance by the scaling
$\rr_i\to \lambda \rr_i$
\cite{another_point},
and solves the effective 2D Schr\"odinger equation
\be
\label{eq:S2D}
E F(R) = -\frac{\hbar^2}{2\bar{m}} \left(\partial_R^2 +\frac{1}{R}\partial_R\right)F(R)+\frac{\hbar^2s^2}{2\bar{m}R^2} F(R).
\ee
The quantity $s^2$ is given by the hyperangular eigenvalue problem. It belongs to a infinite discrete set
and is real since there is no Efimov effect on the unit sphere ($R\neq 0$), 
that is here no three-body Efimov effect.

Mathematically, Eq.(\ref{eq:S2D}) admits for all energies $E$
two linearly independent solutions, respectively behaving as $R^{\pm s}$ for $R\to 0$.
If $s^2>0$, one imposes
$F(R)\sim R^s$, with $s>0$, which is correct except for accidental, non-universal four-body resonances 
(see note [43] in \cite{Werner2006}),
and Eq.(\ref{eq:S2D}) then does not support any bound state. On the contrary, if $s^2<0$, in which case
we set $s=iS$, $S>0$, $F$ experiences an effective four-body attraction, with a fall to the center leading to a unphysical
continuous spectrum of bound states \cite{Morse}.
To make the model self-adjoint, 
one then imposes an extra contact condition \cite{Morse},  as in the usual three-body Efimov case \cite{three_body_parameter}:
\be
\label{eq:Rf}
F(R) \underset{R\to 0}{\sim} \mbox{Im}\, \left[\left(\frac{R}{R_f}\right)^{iS}\right],
\ee
where the four-body parameter $R_f$ depends on the microscopic details of the true, finite range 
interaction \cite{narrow_resonance}.
With the extra condition Eq.(\ref{eq:Rf}) one then obtains from Eq.(\ref{eq:S2D}) 
an Efimov spectrum of tetramers:
\be
\label{eq:En}
E_n = -\frac{2\hbar^2}{\bar{m} R_f^2} e^{\frac{2}{S}\arg \Gamma(1+iS)}e^{-2\pi n/S}, \ \ \ \ \ \forall n \in \mathbb{Z}.
\ee

The whole issue is thus to determine the values of the exponents $s$. In particular, the critical mass ratio  $\alpha_c(3;1)$
corresponds to one of the exponents being equal to zero, the other ones remaining positive. To this end, 
we calculate the zero energy four-body wavefunction with no specific boundary condition on $F(R)$. Then, from Eq.(\ref{eq:S2D}) with $E=0$, 
it appears that $F(R)\propto R^{\pm s}$. The calculation is done in momentum space, with the ansatz
for the Fourier transform of the four-body wavefunction:
\begin{multline}
\label{eq:ansatz_k}
\tilde{\psi}(\kk_1,\kk_2,\kk_3,\kk_4) = \frac{\delta(\sum_{i=1}^{4}\kk_i)}{\sum_{i=1}^{4} \frac{\hbar^2 k_i^2}{2 m_i}} \\
\times \left[D(\kk_2,\kk_3) + D(\kk_3,\kk_1) + D(\kk_1,\kk_2)\right],
\end{multline}
where the fermionic symmetry imposes $D(\kk_2,\kk_1)=-D(\kk_1,\kk_2)$, and the denominator originates from the action
of $H$ in Eq.(\ref{eq:hamil}) written in momentum space. When $H$ acts on one of the three $1/|\rr_4-\rr_i|$ singularities
in Eq.(\ref{eq:BP}),
this produces in the right hand side of Schr\"odinger's equation 
a Dirac distribution $\delta(\rr_4-\rr_i)$ multiplied by a translationally invariant function of
the three fermionic positions, which after Fourier transform gives each of the $D[(\kk_j)_{j\neq i,4}]$ terms
in Eq.(\ref{eq:ansatz_k}).
Taking the Fourier transform of Eq.(\ref{eq:separe}) with $F(R)\propto R^{\pm s}$, and using a power-counting argument,
one finds the scaling law
\be
\label{eq:scaD}
D(\lambda \kk_1,\lambda \kk_2) = \lambda^{-(\pm s+7/2)} D(\kk_1,\kk_2).
\ee
Implementing in momentum space the contact conditions, that is the fact that $O(|\rr_i-\rr_4|)$ vanishes for $\rr_i=\rr_4$ in Eq.(\ref{eq:BP}),
gives rise to an integral equation:
\begin{multline}
\label{eq:minlos}
0=\left[\frac{1+2\alpha}{(1+\alpha)^2}(k_1^2+k_2^2) +\frac{2\alpha}{(1+\alpha)^2}
\kk_1\cdot\kk_2\right]^{1/2} D(\kk_1,\kk_2) \\
+ \int \frac{d^3k_3}{2\pi^2} \frac{D(\kk_1,\kk_3)+D(\kk_3,\kk_2)}
{ k_1^2+k_2^2+k_3^2+
\frac{2\alpha}{1+\alpha}
(\kk_1\cdot \kk_2 + \kk_1\cdot \kk_3 + \kk_2\cdot\kk_3)},
\end{multline}
where we recall that $\alpha=M/m$.
Eq.(\ref{eq:minlos}) can also be obtained as the zero range limit of finite range models \cite{CRAS}.

We now use rotational invariance to impose the value $l\in \mathbb{N}$ of the total angular momentum
of the four-body state and to restrict to a zero angular momentum 
along the quantization axis $z$. Then, according to Eq.(\ref{eq:ansatz_k}), the effective two-body function
$D(\kk_1,\kk_2)$ has the same angular momentum $l$.
This allows to express $D$ in terms of $2l+1$ unknown functions $f^{(l)}_{m_l}$
of three real variables only, the moduli $k_1$ and $k_2$ and the angle $\theta\in [0,\pi]$
between $\kk_1$ and $\kk_2$,
with the fermionic symmetry imposing $f^{(l)}_{m_l}(k_2,k_1,\theta)=(-1)^{l+1} f^{(l)}_{-m_l}(k_1,k_2,\theta)$
\cite{CRAS}:
\be
\label{eq:ansatz_general}
D(\kk_1,\kk_2)=
\sum_{m_l=-l}^{l} \left[Y_l^{m_l} (\gamma,\delta)\right]^* e^{i m_l\theta/2}
f^{(l)}_{m_l}(k_1,k_2,\theta).
\ee
Here $Y_l^{m_l}(\gamma,\delta)$ are the usual spherical harmonics, $\gamma$ and $\delta$ are the polar and azimuthal
angles of the unit vector vector $\mathbf{e}_z$ along $z$ in the direct orthonormal basis
$(\mathbf{e}_1, \mathbf{e}_{2\perp}, \mathbf{e}_{12})$,
with $\mathbf{e}_1=\kk_1/k_1,\mathbf{e}_2=\kk_2/k_2$, $\mathbf{e}_{2\perp}= (\mathbf{e}_2-\mathbf{e}_1\cos\theta)
/\sin\theta$ and $\mathbf{e}_{12}=\mathbf{e}_1\wedge\mathbf{e}_2/\sin\theta$ 
\cite{thus_has}.
The action of parity $\kk_i\to -\kk_i$ on this general ansatz is to multiply each term of index $m_l$
in Eq.(\ref{eq:ansatz_general}) by a factor $(-1)^{m_l}$, which allows to decouple the even $m_l$ terms
(even parity) from the odd $m_l$ terms (odd parity). A relevant example, as we shall see, is the even
parity channel with $l=1$, where the ansatz reduces to a single term,
which is obviously the component along $z$ of a vectorial spinor:
\be
D(\kk_1,\kk_2) \propto \mathbf{e}_z\cdot \frac{\kk_1\wedge\kk_2}{||\kk_1\wedge\kk_2||} f^{(1)}_0(k_1,k_2,\theta).
\ee

The last step is to use the scaling invariance of $D$, see Eq.(\ref{eq:scaD}), setting
\be
f^{(l)}_{m_l}(k_1,k_2,\theta) = (k_1^2 + k_2^2)^{-(s+7/2)/2} (\cosh x)^{3/2} \Phi^{(l)}_{m_l}(x,u)
\ee
where $u=\cos\theta$. The introduction of the logarithmic change of variable $x=\ln(k_2/k_1)$ is motivated by Efimov physics,
and the factor involving the hyperbolic cosine ensures that the final integral equation involves a Hermitian
operator. The fermionic symmetry imposes
\be
\Phi^{(l)}_{m_l}(-x,u)= (-1)^{l+1} \Phi^{(l)}_{-m_l}(x,u)
\ee
which allows to restrict the unknown functions $\Phi^{(l)}_{m_l}$ to $x\geq 0$.
Restricting to $s=iS$, $S\geq 0$, we finally obtain
\begin{multline}
\label{eq:int_eq}
0 = \left[\frac{1+2\alpha}{(1+\alpha)^2} +\frac{\alpha u}{(1+\alpha)^2 \cosh x}\right]^{1/2} \Phi^{(l)}_{m_l}(x,u)  \\
+\int_{\mathbb{R}^+}\! dx' \int_{-1}^{1}\! du' \sum_{m_l'=-l}^{l}
\mathcal{K}_{m_l,m_l'}^{(l)}(x,u;x',u')
\Phi^{(l)}_{m'_l}(x',u').
\end{multline}
The symmetrized kernel
$\mathcal{K}_{m_l,m'_l}^{(l)}(x,u;x',u') =
\sum_{\epsilon,\epsilon'=\pm 1}
(\epsilon \epsilon')^{l+1} K^{(l)}_{\epsilon m_l,\epsilon' m'_l}
(\epsilon x,u;\epsilon' x',u')$
is expressed in terms of the non-symmetrized one given by:
\begin{multline}
K^{(l)}_{m_l,m_l'}(x,u;x',u')=
\frac{\left[(1+\lambda^2)/(1+\lambda'^{2})\right]^{iS/2}(\lambda\lambda')^{3/2}}{[(1+\lambda^2)(1+\lambda'^{2})]^{1/4}} \\
\times 
\int_0^{2\pi} \frac{d\phi}{2\pi^2}
\frac{e^{-im_l\theta/2}\,\langle l,m_l| e^{i\phi L_x/\hbar} |l,m_l'\rangle\, e^{im_l'\theta'/2}}
{1+\lambda^2+\lambda'^{2}+\frac{2\alpha}{1+\alpha} [\lambda u + \lambda' u' +\lambda\lambda'\mathcal{D}]}.
\end{multline}
Here the notation $\mathcal{D}$ in the denominator stands for 
$\mathcal{D}=u u'
+\cos\phi \sqrt{1-u^2} \sqrt{1-u^{'2}}$,
$\lambda=e^x, \lambda'=e^{x'}$,  $L_x$ is the angular momentum operator along $x$, 
$|l,m_l\rangle$ is of spin $l$ and angular momentum $m_l \hbar$ along $z$, and $\phi$ stands for the azimuthal angle of the
vector $\kk_3$ of Eq.(\ref{eq:minlos}) 
in the spherical coordinates related to the basis $(\mathbf{e}_{2\perp},\mathbf{e}_{12},\mathbf{e}_1)$
\cite{matrix_elem}.

\begin{figure}[tb]
\includegraphics[width=8cm,clip=]{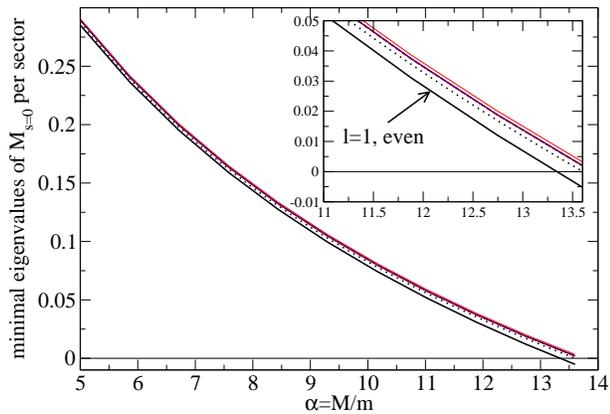}
\caption{Minimal eigenvalues of the Hermitian operator $M_{s=0}$ in each sector of
fixed parity and angular momentum $l$, $0\leq l\leq 6$, as functions of the mass ratio $\alpha=M/m$.
Only the curve for the even sector of $l=1$ crosses zero for $\alpha< 13.607$, corresponding to the occurrence
of a four-body Efimov effect in that sector. The other curves all remain above zero. They strongly overlap
and are barely distinguishable at the scale of the figure.
The dotted line is the analytical prediction $\Lambda(k=0,\alpha)$ for the lower border of the continuum
in the spectrum of $M_{s=0}$.  The inset is a magnification.
In the numerics, $x$ and $u$ were discretized with a step $dx=du=1/10$, and $x$ was truncated
to $x_{\rm max}=20$.}
\label{fig:val_min}
\end{figure}

We first look for the critical mass ratio for the 3+1 fermionic problem $\alpha_c(3;1)$, 
which is the minimal value of $\alpha$ such that the integral
equation Eq.(\ref{eq:int_eq}) is satisfied for $S=0$. Rewriting Eq.(\ref{eq:int_eq}) as $0=M_s [\Phi]$, where
$M_s$ is a Hermitian operator, we calculated numerically the minimal eigenvalues of $M_{s=0}$ as functions of the mass ratio 
$\alpha$, within each subspace of fixed parity and angular momentum $l$, $0\leq l\leq 6$. As shown in Fig.\ref{fig:val_min}, 
such a minimal eigenvalue vanishes for $\alpha<13.607$ only in the even sector of angular momentum $l=1$.
We also unfruitfully explored $l=7, 8, 9, 10$. 
We thus find that the four-body Efimov effect takes place only in the even sector of $l=1$, and sets in above a critical mass ratio
\cite{gain_precision}
\be
\alpha_c(3;1) \simeq 13.384,
\ee
quite close to the $2+1$ critical value $\alpha_c(2;1)\simeq 13.607$.

To gain some insight on this result,
we have studied analytically an important feature of the spectrum of $M_{s=0}$, 
the lower border of its continuum. When $x, x'\to +\infty$, which corresponds physically to having $k_2\gg k_1$
in the function $D(\kk_1,\kk_2)$,
both the symmetrized and non-symmetrized kernels reduce to the asymptotic form
\begin{multline}
\mathcal{K}_{m_l,m'_l}^{(l)}(x,u;x',u') \sim 
e^{iS(x-x')}e^{-im_l\theta/2} e^{im_l'\theta'/2} \\
\times \int_0^{2\pi}\frac{d\phi}{4\pi^2} 
\frac{\langle l,m_l| e^{i\phi L_x/\hbar} |l,m_l'\rangle}
{\cosh(x-x') + \frac{\alpha}{1+\alpha} \mathcal{D}}.
\end{multline}
Since $\mathcal{D}$ is independent of $x$ and $x'$,
this is invariant by translation over the $x$ coordinates, leading to a continuous spectrum of asymptotic plane wave
eigenfunctions.
In the even sector of angular momentum $l=1$, we found that $\Phi^{(1)}_0(x,u) \sim e^{ikx}\sqrt{1-u^2}$ 
gives rise to an eigenfunction in the continuous spectrum of $M_{iS}$ with the real eigenvalue 
$\Lambda(k-S,\alpha)$ \cite{rapidly}
where 
\be
\label{eq:Lambda}
\Lambda(k,\alpha) = \cos2\beta + \frac{(1-ik)\sin[2\beta(1+ik)]-\mbox{c.c.}}{2(1+k^2)\sin^22\beta \sin(i k \pi/2)}.
\ee
In Eq.(\ref{eq:Lambda})  we have set for convenience $\sin 2\beta =\alpha/(1+\alpha)$ with $\beta \in [0,\pi/2[$. 
For real $k$, this function $\Lambda(k,\alpha)$ has a global minimum in $k=0$.
We expect that $\Lambda(k=0,\alpha)$ is the 
lower border of the continuous spectrum of $M_{s=0}$.  Since $\Lambda(0,\alpha)$ exactly vanishes
for the three-body critical mass ratio $\alpha_c(2;1)\simeq 13.607$, our asymptotic analysis amounts
to uncovering the three-body problem as a limit $k_2/k_1\to +\infty$ of the four-body problem.

We tested this prediction against the numerics, plotting
in Fig.\ref{fig:val_min} the quantity $\Lambda(k=0,\alpha)$ as a function of $\alpha$ in dotted line. Except for the even sector of $l=1$,
the minimal numerical eigenvalues are close to $\Lambda(k=0,\alpha)$; the fact that they are slightly above is due to
a finite $x_{\rm max}$ truncation effect, that indeed decreases for increasing $x_{\rm max}$ (not shown).
This implies that the eigenfunctions corresponding to these minimal eigenvalues are extended, that is not square
integrable.  The numerics agrees with this analysis.
In the even sector of $l=1$, the minimal numerical eigenvalue is clearly below $\Lambda(0,\alpha)$, for all values
of $\alpha$ in Fig.\ref{fig:val_min}. This indicates that the corresponding eigenvector must
be a bound state of $M_{s=0}$, with a square integrable eigenfunction $\Phi^{(1)}_0(x,u)$. This is confirmed
by the numerics, which shows that at large $x$, 
$\Phi^{(1)}_0(x,u)\propto \sqrt{1-u^2} e^{-\kappa x}$ \cite{weakly}.
The analytical reasoning even predicts the link between the minimal eigenvalue $\Lambda_{\rm min}$ of $M_{s=0}$ and the decay
constant $\kappa$: The plane wave $e^{ik x}$ is analytically continuated into a decreasing
exponential if one sets $k=i\kappa$, so that
\be
\label{eq:kappa}
\Lambda_{\rm min} = \Lambda(i\kappa,\alpha).
\ee
Numerically, we have successfully tested this relation for various values of $\alpha$,
and we also found that $M_{s=0}$ has no other bound state in the even sector of $l=1$.

\begin{figure}[tb]
\includegraphics[width=8cm,clip=]{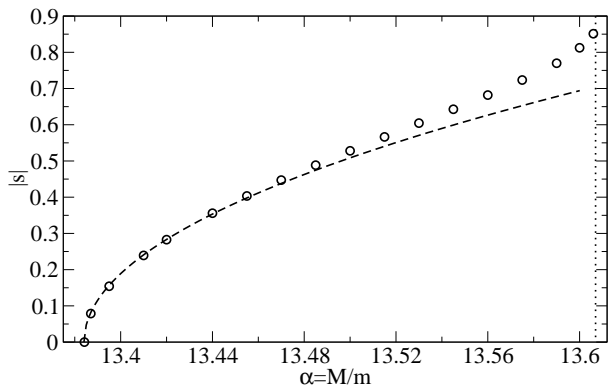}
\caption{In the Efimovian channel $l=1$ with even parity, modulus of the purely imaginary Efimov exponent
$s=iS$ as a function of the mass ratio $\alpha=M/m$. In the numerics, $x_{\rm max}$ ranges from 40 to $120$, $dx=1/10$, $d\theta=\pi/20$.
The dashed line results from a linear fit of $|s|^2$ as a function of $\alpha$ in a vicinity of the critical value
$\alpha_c(3;1)$, $|s|^2_{\rm fit}\simeq 2.23\times(\alpha-\alpha_c)$. The vertical dotted line indicates the $2+1$
critical value $\alpha_c(2;1)$.}
\label{fig:S}
\end{figure}

Finally, we completed our study of the four-body Efimov effect by calculating, as a function of the mass ratio $\alpha$,
the exponent $s=iS$ in the even sector of $l=1$, the real quantity $S$ being such that
the operator $M_{iS}$ has a zero eigenvalue. The result is shown in Fig.\ref{fig:S}.
Close to the $2+1$ critical mass ratio $\alpha_c(2;1)\simeq 13.607$, the values of $|s|$ are 
not far from the three-boson Efimov exponent $|s_0|\simeq 1$ proved to have observable effects \cite{manips_Efimov}.
Close to the $3+1$ critical mass ratio $\alpha_c(3;1)$, $|s|$ varies as expected as $(\alpha-\alpha_c)^{1/2}$ (see dashed line).
Low values of $|s|$ may lead to extremely low Efimov tetramer binding energies:
For an interaction of finite range $b$, setting $R_f\approx b$ and $n=1$ in Eq.(\ref{eq:En}),
we estimate the ground state Efimov tetramer energy for $|s|\ll 1$ as
$E_{\rm min}^{\rm Efim} \approx -e^{-2\pi/|s|} \hbar^2/(2\bar{m} b^2)$ \cite{autre_cond}.
For $|s|=0.5$, taking the mass of ${}^3$He for $m$ and a few nm for $b$ gives
$E_{\rm min}^{\rm Efim}/k_B$ in the nK range, accessible to cold atoms.
Moreover, for a large but finite scattering length $a$, successive Efimov
tetramers come in for values of $a$ in geometric progression of ratio $e^{\pi/|s|}$, so that too low values of $|s|$
require unrealistically large values of the scattering length.
Another experimental issue is
the narrowness of the mass interval. Several pairs of atomic species have a mass ratio
in the desired interval, e.g.\ ${}^3$He${}^*$ and ${}^{41}$Ca ($\alpha \simeq 13.58$),
and with exotic species, ${}^{11}$B and ${}^{149}$Sm ($\alpha \simeq 13.53$),
${}^{7}$Li and ${}^{95}$Mo ($\alpha \simeq 13.53$).
A more flexible solution is to start with usual atomic species having a slightly off mass ratio,
such as ${}^{3}$He${}^*$ and ${}^{40}$K ($\alpha\simeq 13.25$),
and to use a weak optical lattice to finely tune the effective mass of one of the species \cite{Petrov_idee}.

To conclude, in the zero range model at unitarity, we studied the interaction
of three same spin state fermions of mass $M$ with another particle of
mass $m$. For $M/m< 13.384$, no Efimov effect was found.
Over the interval $13.384 < M/m < 13.607$, remarkably a purely four-body Efimov effect takes place,
in the sector of even parity and angular momentum $l=1$, that may be observed with a dedicated
cold atom experiment.
For $M/m>13.607$, the three-body Efimov effect sets in, and the zero range model 
has to be supplemented by three-body contact conditions that break its separability.
The intriguing question of wether the Efimov tetramers then survive as resonances, decaying in a trimer plus a free atom,
is left for the future. F.\ Werner is warmly thanked for discussions.


\begin{thebibliography}{99}
\bibitem{narrow_resonance} For a narrow resonance, $R_f$ may be proportional
to the effective range as for the three-body Efimov effect \cite{MoraGogolin}.
\bibitem{matrix_elem} The matrix elements $\langle l,m_l|e^{i\phi L_x/\hbar}|l,m_l'\rangle$ are evaluated by
insertion of a closure relation in the
eigenbasis of $L_x$. One then faces integrals $J_n = \int_0^{2\pi} d\phi \frac{e^{in\phi}}{b_0+b_1\cos\phi}
=2\pi z_0^{|n|}/\sqrt{b_0^2-b_1^2}$ with $z_0=-(b_0/b_1)+\sqrt{(b_0/b_1)^2-1}.$
\bibitem{gain_precision} To gain in precision, we used  $\theta$ rather than $u=\cos\theta$ as a variable,
with $x_{\rm max}$ up to $40$, $dx=d\theta/\pi$ down to $1/40$.
Taking as unknown functions $(x,\theta)\to \sqrt{\sin\theta}\, \Phi^{(l)}_{m_l}(x,\cos\theta)$ preserves
the operator hermiticity.
\bibitem{rapidly} This is more rapidly obtained by taking the large $k_2$ limit directly in Eq.(\ref{eq:minlos})
and inserting the ansatz $D(\kk_1,\kk_2) \sim \mathbf{e}_1\wedge \mathbf{e}_2 k_2^{-7/2} (k_2/k_1)^{ik+3/2}$.
\bibitem{weakly} The corresponding eigenvalue of $M_{s=0}$ is thus weakly sensitive to truncation effects.
\end{thebibliography}
\end{document}